# On the rotational alignment of graphene domains grown on Ge(110) and Ge(111)

P. C. Rogge[1,2], M. E. Foster[3], J. M. Wofford[1,2*], K. F. McCarty[3], N. C. Bartelt[3], and O. D. Dubon[1,2]
Affiliations
1. Department of Materials Science and Engineering, University of California, Berkeley, Berkeley, CA 94720
2. Materials Sciences Division, Lawrence Berkeley National Laboratory, Berkeley, CA 94720
3. Sandia National Laboratories, Livermore, CA 94550

Corresponding author: P. C. Rogge: progge@berkeley.edu
*current affiliation: Paul-Drude-Institut für Festkörperelektronik, Berlin, Germany 10117

**Abstract**
We have used low-energy electron diffraction and microscopy to compare the growth of graphene on hydrogen-free Ge(111) and Ge(110) from an atomic carbon flux. Growth on Ge(110) leads to significantly better rotational alignment of graphene domains with the substrate. To explain the poor rotational alignment on Ge(111), we have investigated experimentally and theoretically how the adatom reconstructions on Ge interact with graphene. We find that the ordering transition of the adatom reconstruction of Ge(111) is not significantly perturbed by graphene. Density functional theory calculations show that graphene on reconstructed Ge(110) has large-amplitude corrugations, whereas it is remarkably flat on reconstructed Ge(111). We argue that the absence of corrugations prevents graphene islands from locking into a preferred orientation.

# 1. Introduction

Because of its intrinsic two-dimensional character, graphene interacts weakly with the substrates on which it is grown. Consequently, graphene domains tend to nucleate in random orientations, producing polycrystalline films [1-5]. Much has been learnt about the factors that control domain orientation, particularly on metal substrates. For example, it has been shown that a single, energetically preferred orientation exists on Ir(111), where the preferred graphene orientation is dictated by the amplitude of the moiré corrugation of the graphene film induced by film-substrate interactions [6]. Recently, germanium, a group IV semiconductor, has emerged as a viable substrate for graphene epitaxy [7-9]. High quality epitaxial growth on semiconductor wafers is very attractive because it presents a path toward production by existing very large scale integration processes. Graphene growth on Ge(110) by chemical vapor deposition (CVD) has been shown to exhibit considerably improved rotational alignment compared to growth on Ge(111) [7]. Here, we clarify the reasons for this improvement.

Due to the directional nature of covalent bonds, semiconductor surfaces undergo distinctive reconstructions to minimize dangling bonds, which presents a heterogeneous bonding environment for graphene in contrast to closed-packed metal surfaces. In the case of germanium, the surface reconstructions consist of ordered adatom phases that exhibit first-order disordering phase transitions with temperature [10, 11]. The high-temperature phases are believed to be disordered versions of the low-temperature phases such that both contain similar densities of adatoms [10]. It is critical to understand how graphene films interact with this complex growth substrate in order to optimize growth on germanium and other covalently bonded materials.

One way to probe the interaction between Ge and graphene, and thus to understand what dictates the degree of alignment, is to determine how the presence of graphene affects the Ge surface reconstruction—that is, how graphene affects the nature of the ordered or disordered adatom phases and the transition between them. This effect is a measure of how strongly the graphene is coupled to the substrate. To probe this effect, we have used low-energy electron microscopy (LEEM) and low-energy electron diffraction (LEED) to study graphene growth on Ge(111) and Ge(110) while monitoring the nature of the surface reconstruction. We find that the clean Ge(111) surface reconstruction remains even when supporting a graphene film, whereas the Ge(110) surface reconstruction is more strongly affected. Further, we used density functional theory (DFT) to investigate the influence of the Ge surface reconstruction on the structure of the graphene film. We find that graphene on the adatom-reconstructed Ge(111) surface is remarkably flat relative to graphene on Ge(110), which we argue (following Ref. 6) explains the improved rotational order on Ge(110).

# 2. Methods

Germanium substrates were cut from single crystal wafers and prepared in the low-energy electron microscope by cycles of Ar-ion sputtering (1.5 keV, $5\times10^{-6}$ torr Ar, 5 minutes, base pressure $1\times10^{-10}$ torr) and annealing at ~800 °C. Temperature was measured by a W-Rh thermocouple in contact with a tantalum disc supporting the Ge sample. Temperature readings were calibrated using the well-known transition temperatures of the surface reconstructions for Ge(111) [10] and Ge(110) [11]. In general, the thermocouple reading was ~50 K less than the estimated actual sample temperature.

Heating a sample until melting further confirmed this estimate. The temperatures reported here are adjusted and are estimated to be accurate to ±20 K. Carbon was deposited by physical vapor deposition (PVD), where an elemental carbon vapor flux was produced by heating a graphite rod by bombardment with 3 keV electrons. After growth and *in-situ* characterization, samples were removed from the microscope and exposed to the ambient atmosphere for at least 24 hours. Samples were then placed in room-temperature de-ionized water for 24 hours. A silicon wafer with a thick oxide layer was used to collect graphene floating on the water surface. The wafer was then dried by heating on a hot plate at 150 °C for 10 minutes. Raman spectroscopy was used to evaluate the transferred graphene.

For all DFT calculations, the non-local optB86b-vdW exchange-correlation functional [12] was used within the unrestricted Kohn-Sham formalism. The projector-augmented-wave (PAW) method, as implemented in VASP [13-16], was utilized to model the core electrons and the wavefunctions were expanded in a plane-wave basis with an energy cutoff of 400 eV. A gamma-point sampling of the Brillouin zone was used for all calculations except bulk Ge where a 10x10x10 k-point grid was used. The optB86b-vdW exchange-correlation functional was used for its ability to approximately account for dispersion interactions (van der Waals forces), which is imperative to qualitatively model the intermolecular interactions between graphene and metal surfaces. Moreover, it has been previously demonstrated that the optB86b-vdW functional is currently among the most accurate vdW functionals [12, 17, 18].

## 3. Results

LEEM images were acquired in mirror-mode, where the incident electron energy (~1 eV used here) is less than the surface work function. In this mode, image contrast is sensitive to surface defects and changes in work function [19, 20]. After Ar-ion sputtering, the surface is featureless except for a low density of defects as seen in Fig. 1(a). Observable defects with diameter ~700 nm were present in low density (~2 x $10^4$ cm$^{-2}$) as well as smaller defects, ~200 nm in diameter, with higher density (~4 x $10^5$ cm$^{-2}$). The size and density of these defects increased with increasing sputtering cycles.

### 3.1 Graphene on Ge(111)

In general, LEED indicated a high-quality Ge(111) surface after sputtering as evidenced by the well-known c(2x8) reconstruction (Fig. 1(b)). The c(2x8) reconstruction is an ordered adatom phase with three equivalent domain orientations that result in the four eighth-order satellite diffraction spots as labeled in Fig. 1(c). The c(2x8) reconstruction exhibits a first-order transition at 300 °C to a disordered (1x1) phase [10].

After surface preparation, the sample was heated and carbon was deposited. The LEEM image in Fig. 1(d) shows the surface after depositing carbon for 80 minutes at 890 °C. The surface contrast is clearly different from the pristine surface in Fig. 1(a). LEED obtained at high temperature (Fig. 1(e)) indicates the presence of a new diffraction ring with the spacing expected for graphene. Auger electron spectroscopy confirmed the presence of C on the surface after deposition (not shown). As shown in Fig. 1(g), the changes in LEED intensity along the ring are uniform with random variations. This suggests that the graphene islands nucleate in random orientations with equal probability on Ge(111). The illumination diameter for this area is 5 μm.

Reducing the illuminated area to 1.5 µm in diameter results in the discrete graphene LEED spots shown in Fig. 1(f). These distinct spots indicate that graphene domains with high crystalline quality exist. By counting the LEED spots and comparing their relative intensity, we estimate roughly that the aperture illuminated 10-20 graphene domains (assuming that the domains are of comparable size). Assuming a complete graphene layer (discussed below) gives an average grain size of 100-500 nm, which is consistent with previous reports of growth by chemical vapor deposition (CVD) [7].

We were unable to effectively image the growth of individual graphene domains in real-time. LEEM requires a high voltage (15 kV here) between the sample and the objective lens. Because growth is conducted at temperatures near the Ge melting point, any arcing caused by this high voltage often resulted in the destruction of the sample. Furthermore, imaging growth is complicated by the small graphene domain size and the low contrast between pristine Ge and graphene-covered Ge.

With a graphene layer covering the surface, it is not unreasonable to expect that it would significantly affect the c(2x8) reconstruction. Surprisingly, this is not the case. Upon cooling, the c(2x8) reconstruction forms as seen in the LEED pattern in Fig. 2(a). Indeed, as one sees in Fig. 2(b), the transition temperature of the graphene-covered surface is nominally only 10 K higher than pristine Ge(111). This is within the error of the temperature measurement. We suggest two possible explanations for the unchanged transition temperature. The first is that the transition we observe occurs only in regions not covered by graphene. The second is that the relative free energies of the ordered and disordered phases are not affected significantly by the presence of graphene. We establish that the latter is true by showing that the disordered phase is different from that of the pristine surface over the entire sample.

The average intensity of the four 1/8-order c(2x8) spots were measured while cooling at a rate of -0.5 K/sec for the pristine and graphene-covered samples, and are compared in Fig. 2(c) as a function of the LEED spot intensity change. Because of the transition temperature shift noted above, the change in intensity is used as a common metric, where the denoted intensity change is the percentage of the difference between the initial and final intensity of the c(2x8) LEED spots (Fig. 2(b)). From the LEED images in Fig. 2(c), one sees that, consistent with Ref. 10, the 1/2-order spot broadens significantly above the transition temperature for the pristine surface. This is confirmed by plotting their intensity along the red, dashed lines as the transition occurs (Fig. 2(d)). The maximum intensity of the 1/2-order spot decreases as it broadens. In contrast, the graphene-covered sample in Fig. 2(e) shows no broadening. Here, the maximum intensity begins decreasing at a point further along the transition. For the pristine surface, the intensity of the 1/2- and 1/8-order spots monotonically decreases as the phase transition proceeds. Comparing to the graphene-covered surface in Fig. 2(e), the 1/2-order spot maintains its initial intensity while the 1/8-order spots decrease in intensity by 50%. In Ref. 10, the broadening of the ½-order spots is interpreted in terms of short-range order in the positions of domain walls between ordered regions. Evidently graphene decreases this ordering. The measured differences in the diffraction pattern are independent of the LEED sampling location. We did not detect the broadening associated with the pristine surface anywhere on the graphene-covered sample. Thus, we conclude that the entire surface was covered by at least one monolayer of graphene. That the transition temperature is unchanged suggests that the domain ordering at high temperature has negligible effect on the free energy of the disordered phase.

### 3.2 Graphene on Ge(110)

Similar surface preparation and imaging conditions were used for Ge(110). The post-sputtered surface is shown in Fig. 3(a) along with its corresponding LEED pattern in Fig. 3(b). Compared to Ge(111), the Ge(110) surface is more complex. Multiple features have been reported, including uniformly spaced terraces forming {17 15 1} facets, a c(8x10) reconstruction, and a (16x2) reconstruction [11, 21, 22]. These features are stable in specific temperature windows and their formation can be kinetically limited [21]. The LEED pattern in Fig. 3(b) has spots with half the periodicity of the first order spots (red circles) in the [001] direction and one-eighth the periodicity in the [111] direction (see Fig. 3(d) for reference), which have been previously attributed to {17 15 1} facets [11]. We note that these facets may occur in the [779] direction as well; however, facets oriented along a single direction, as seen here, have been previously reported for miscuts less than 2° [11].

Fig. 3(c) shows the surface after depositing C for 45 minutes at 870 °C. The surface exhibits similar features as those seen on graphene-covered Ge(111), and the presence of graphene is confirmed by the LEED pattern in Fig. 3(d). The graphene exhibits significantly improved alignment compared to graphene on Ge(111). As shown in Fig. 3(e), diffraction from the graphene does not result in a ring but is localized to within 10° of a preferred orientation. This preferred orientation is along the Ge[1$\bar{1}$0] direction of the Ge(110) surface, that is, the graphene lattice vectors align with Ge[1$\bar{1}$0]. Here, an illumination aperture with a 10 μm diameter was used, and the LEED pattern did not change with decreasing aperture size. The same preferred alignment was seen as the illumination region was scanned across the surface. Previous work has suggested that the hydrogen termination of Ge induces graphene alignment on Ge(110) [7]. Because our growth by PVD occurs in the absence of hydrogen, our results show that hydrogen is not responsible for the preferred graphene domain alignment on the (110) surface.

In contrast to Ge(111), the graphene-covered surface is characterized by a significantly different LEED pattern (Fig 3(f)) than that of the pristine surface upon cooling. The {17 15 1} facets are nearly extinguished—the faint appearance of features misoriented relative to the facets seen in Fig. 3(b) are attributed to the facets in the [779] direction. Clearly the Ge(110) surface is affected by the presence of graphene. The extinguished faceting seen in LEED was observed uniformly across the sample. Thus, we conclude that, at a minimum, a full graphene monolayer is present.

### 3.3 Water liftoff and Raman results

While growth on Ge wafers provides significant advantages for process integration, the graphene film must still be separated from the Ge substrate because Ge is a good electrical conductor at room temperature due to its relatively small band gap—less than 0.7 eV. Previous efforts achieved liftoff from Ge by depositing a thin layer of Au on the graphene and then peeling off the gold-graphene film [7]. Then, the gold film was removed by chemical etching, a process that can result in significant deterioration of graphene's electrical performance [23].

We propose a less invasive transfer technique that utilizes the water-soluble nature of germanium oxide [24]. Fig. 4(a) shows an optical microscopy image of fragments collected after soaking a sample of graphene grown on Ge(110) in water. The micro-Raman spectra shown in Fig. 4(b) were taken from the three spots labeled in Fig. 4(a). The Raman spectra exhibit strong D and G peaks, and small 2G peaks, which is consistent with defective graphene.

It is likely that the underlying germanium surface oxidized after removal from the ultra-high vacuum LEEM environment, especially given the defective nature of the film (which may facilitate oxygen penetration). Because germanium oxide is water soluble, we speculate that the surface oxide dissolved during water soaking, resulting in the release of the graphene film. This technique is supported by previous efforts (not shown here), where multilayer graphene films of millimeter dimensions were transferred using only water. We expect monolayer films to be transferable by optimizing this process.

The graphene we have grown on Ge(110) has a small spread of angles (10 degrees, Fig. 3(e)), causing grain boundaries in the completed graphene film. Due to the small grain size, the grain boundary density is likely high and can account for the low quality detected by Raman spectroscopy.

### 3.4 DFT calculations of the structure of graphene on Ge(110) and Ge(111)

To investigate the differences in the alignment, we have calculated the structure of graphene on the two substrates by DFT. Our focus is on the role of the adatom structure in graphene binding and morphology. We start by considering the ordered low-temperature structures as they have been well characterized. For the Ge(110) surface, we studied the c(8x10) reconstruction for computational convenience: the unit cell size best matches graphene. Since the adatom structure of the c(16x2) reconstruction is similar, we expect similar graphene-adatom interactions.

Reconstructed surfaces of Ge(111)c(2x8) and Ge(110)c(8x10) were constructed by adding adatoms at positions given by previous experimental and theoretical studies [25, 26] to bulk truncated surfaces created from the optimized cubic close-packed structure ($a = b = c = 5.763$ Å). The reconstructed surfaces were then relaxed until all forces were less than 0.02 eV/Å. The graphene sheet was then stretched or compressed over the Ge substrate in an effort to minimize lattice strain. The graphene sheet was stretched 1.87% and compressed 0.65% on the Ge(111)c(2x8) surface in the *a* and *b* directions, respectively. On the Ge(110)c(8x10) surface, a stretch of 3.4% in both directions was required [27]. The resulting optimized structures and pertinent structural information are shown in Fig. 5. The binding and graphene deformation energies per carbon atom of the two systems were determined as follows:

$$E_{BE} = \frac{E_{G/Ge} - E_G - E_{Ge}}{N}, \tag{1}$$

$$E_{def} = \frac{E_{G-def} - E_G}{N}, \tag{2}$$

where $E_{G/Ge}$ is the total energy of the system, $E_G$ is the total energy of the optimized isolated graphene layer (in the respective Ge surface unit cell), $E_{G-def}$ is the total energy of the graphene sheet in the G/Ge geometry, $E_{Ge}$ is the total energy of the Ge surface, and *N* is the number of carbon atoms. The resulting graphene binding energies on the Ge(111)c(2x8) and Ge(110)c(8x10) surfaces are -40 and -37 meV/C, respectively. The relatively weak interaction can be accredited to the Ge adatoms; although the graphene sheet is relatively close to the adatoms (~3 Å), on average the sheet lies 3.7-3.8 Å above the Ge surface atoms in both cases, slightly above the typically van der Waals distance. The predicted graphene deformation energies on the Ge(111)c(2x8) and Ge(110)c(8x10) surfaces are 0.61 and 1.63 meV/C respectively. The difference in energy correlates directly to the difference in the observed corrugation patterns/amplitudes (see Fig. 5(d)).

We attribute the extremely small corrugations of Ge(111) to the fact that graphene is stiff and cannot respond to the relatively short wavelength corrugations corresponding to the adatom-adatom separations (~8 Å) of the c(2x8) reconstruction. At the temperatures of the growth experiments, the reconstructions modeled here are disordered. However, graphene on Ge(111) is likely to remain flat because the average adatom separation is not expected to change significantly. For example, the adatom density on Si(111) is approximately temperature independent despite having a high-temperature "1×1" phase and multiple metastable reconstructions [28].

## 4. Discussion

The domain size and the significantly improved rotational order of graphene on Ge(110) compared to Ge(111) is consistent with previous reports of graphene grown by CVD [7]. However, our results call into question some of the conclusions obtained in Ref. 7. The authors suggest that hydrogen plays a critical role in obtaining the orientation preference. Here, growth occurred in complete absence of hydrogen gas and highly oriented films are still obtained. Instead, our DFT calculations suggest a reason for the poor alignment on Ge(111). For the energy of an incommensurate graphene film to depend on orientation, it must be corrugated [6]. So the lack of corrugation in the DFT calculations for Ge(111) suggests that there is no energetically preferred alignment. The reasons for good alignment on Ge(110) is less clear. While the c(8x10) reconstruction gives a large corrugation of the graphene sheet, it is unclear if this corrugation would persist to high temperature. It is possible that the relatively small number of adatoms on Ge(110) are still grouped pentagons at high temperature, which would result in large average separations between pentagons and allow the graphene lattice to become corrugated with an energy which is sensitive to orientation. But regardless of the corrugation, that the low-temperature surface reconstruction of Ge(110) changes due to graphene's presence suggests a stronger dependence of the interaction with the structure (and thus orientation) of the substrate.

## 5. Conclusion

We find that graphene grown on Ge(110) by PVD has fewer rotational domains than growth on Ge(111), consistent with previous CVD results. The surface underneath graphene for both terminations is reconstructed. The insignificant change in disordering temperature of the Ge(111)c(2x8) adatom reconstruction underneath graphene compared to clean Ge indicates a very weak interaction. This is supported by DFT calculations, which show that graphene is not significantly corrugated by the reconstruction. On the other hand, the reconstruction underneath graphene on Ge(110) is strongly affected by the graphene, indicating a more significant interaction with the adatoms. Indeed, DFT shows that graphene on reconstructed Ge(110) is strongly corrugated. We suggest that the absence of corrugation on reconstructed Ge(111) is responsible for the poor rotational alignment compared to Ge(110).

Acknowledgments

This work was supported by the Director, Office of Science, Office of Basic Energy Sciences, Division of Materials Sciences and Engineering, of the U.S. Department of Energy Contract No. DE-AC04-94AL85000 (SNL) and by the NSF under Grant No. DMR-1105541 (ODD, PCR, JMW).

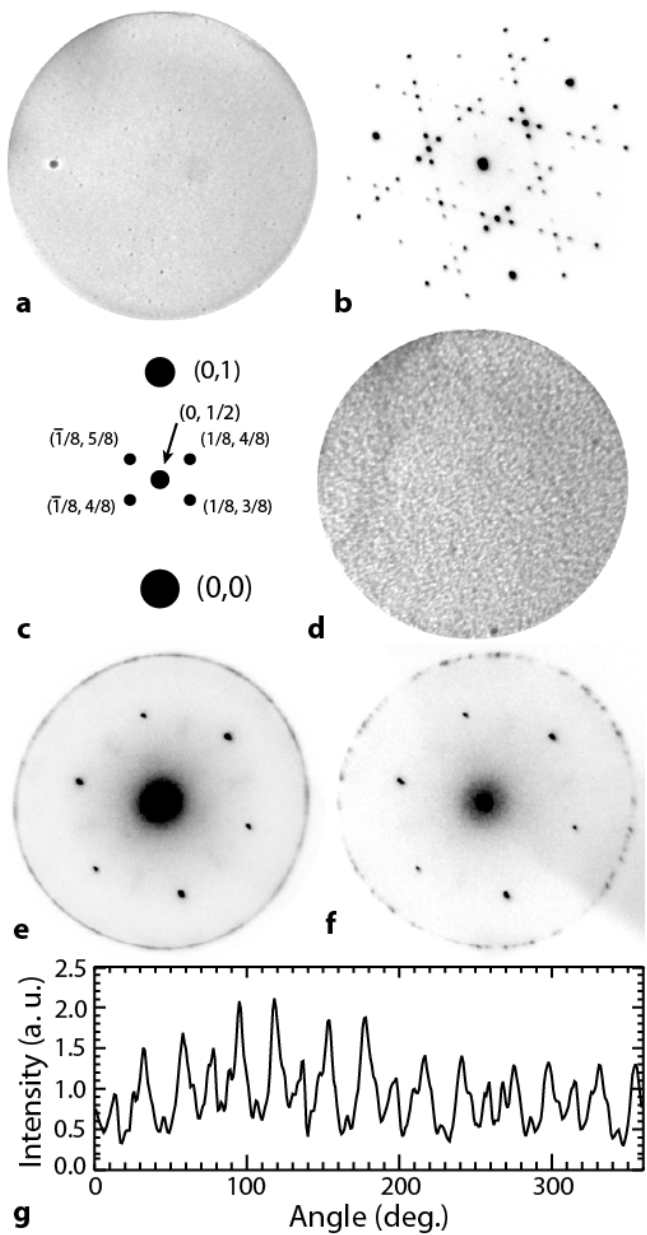

FIG. 1. (a) Initial Ge(111) surface after sputtering, field-of-view (FOV) = 25 μm, T = 25 °C. (b) LEED pattern of initial surface showing c(2x8) reconstruction, electron energy ("start voltage" or SV) = 19.9 eV (c) Schematic of the c(2x8) LEED reconstruction. (d) Ge surface after C deposition at 890 °C, FOV = 25 μm. (e) LEED pattern after C deposition at T = 500 °C showing first-order Ge(111) spots and a graphene diffraction ring, SV = 35.5 eV. (f) LEED pattern with smaller aperture showing distinct graphene diffraction spots, SV = 35.5 eV. (g) Intensity of the graphene LEED pattern extracted from (e) shows random and uniform variations with orientation.

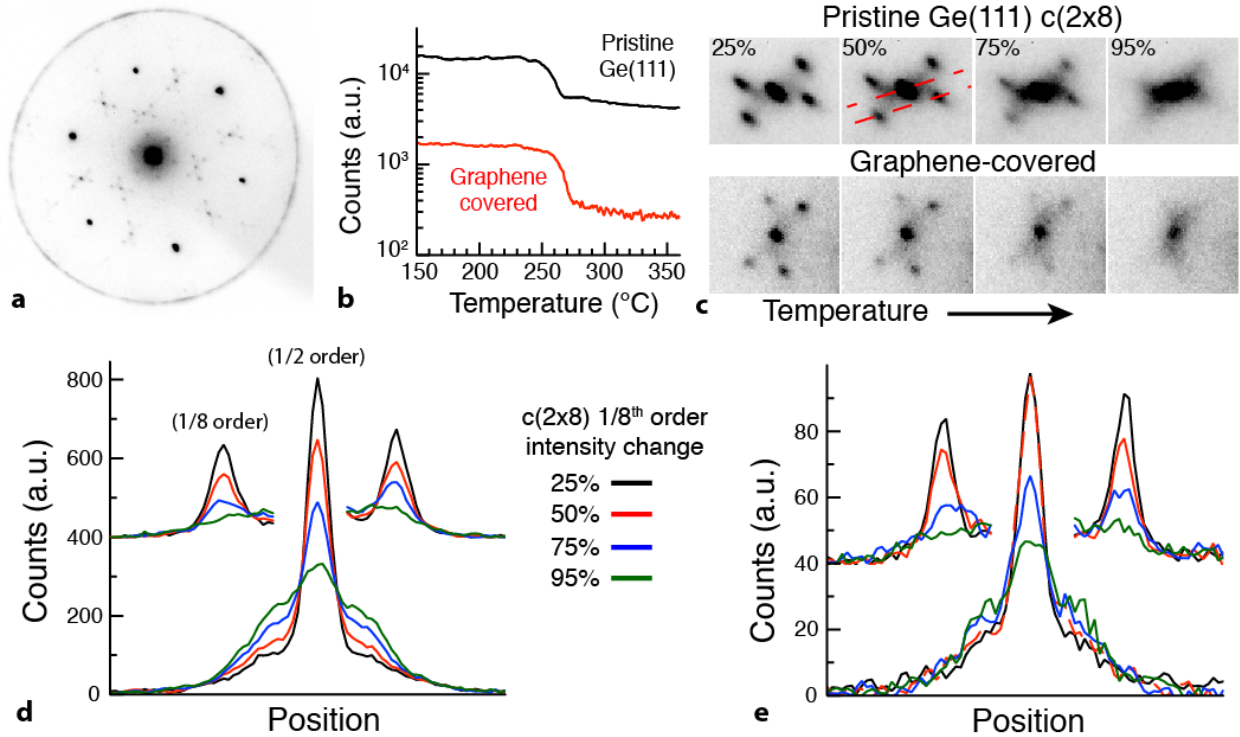

FIG. 2. (a) Low temperature LEED pattern showing presence of both graphene and Ge(111) c(2x8) reconstruction, SV = 36.7 eV. (b) Comparison of the c(2x8) phase transition with temperature for pristine and graphene-covered Ge(111) by monitoring the 1/8th order c(2x8) reconstruction beams. (c) Comparison of the evolution of c(2x8) LEED spots with temperature for both pristine and graphene-covered Ge(111) compared at similar points through their respective changes in c(2x8) intensity, as determined by the intensity changes in (b) and denoted by the percentage value. (d)-(e) Intensity profiles of 1/2 and 1/8-order c(2x8) LEED spots taken along the red dashed lines in (c) and plotted together for pristine Ge(111) c(2x8) reconstruction (e) and the graphene-covered Ge surface (f) at various percentages of intensity change. 1/8-order intensity shifted in y-axis for clarity, scale is maintained.

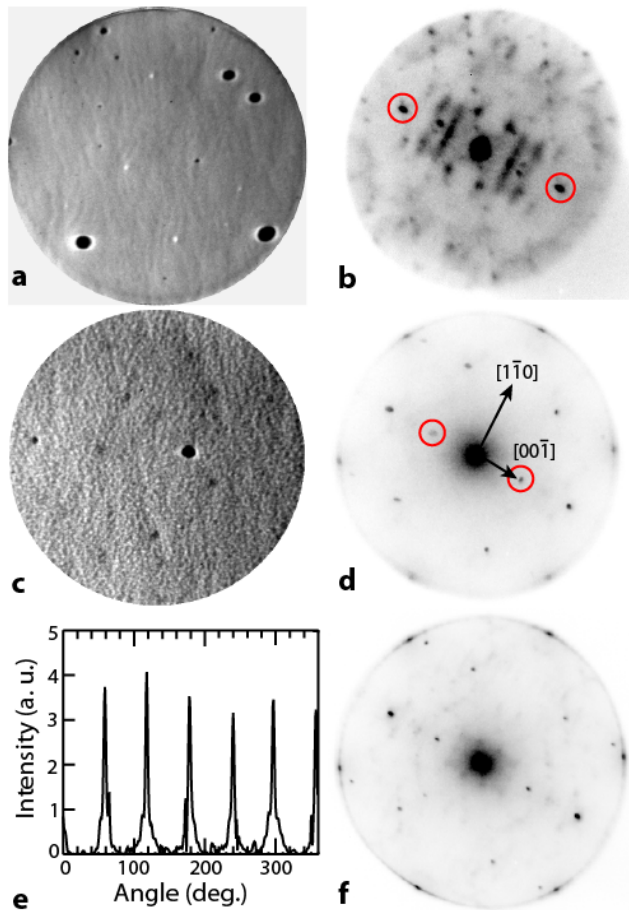

FIG. 3. (a) LEEM image of pristine Ge(110) surface; FOV = 15 μm. (b) LEED pattern of pristine surface. First-order spots highlighted by red circles, superstructure spots attributed to {17 15 1} facets; SV = 11.9 eV. (c) LEEM image after C deposition for 45 minutes at 870 °C; FOV = 15 μm. (d) LEED pattern after C deposition showing localized graphene diffraction spots. For reference, red circles highlight the same spots in (b). T = 870 °C, SV = 38.1 eV. (e) Intensity of the graphene LEED pattern extracted from (d) shows that the graphene is strongly localized to a single orientation. (f) LEED pattern after cooling to 90 °C, SV = 37.3 eV.

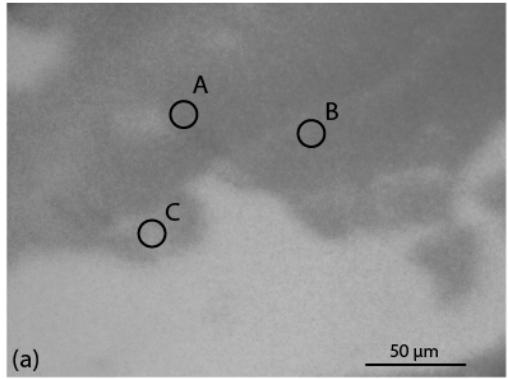

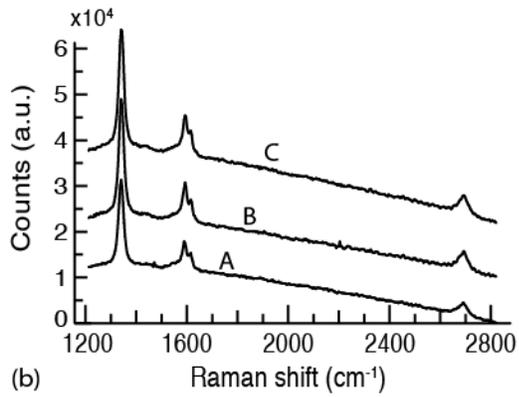

FIG. 4. (a) Optical microscopy image of a graphene flake on SiO$_2$ wafer after soaking a graphene-covered Ge(110) sample in water and collecting the flakes with the wafer. (b) Raman spectra taken at the three labeled locations in (a). The Raman peaks are consistent with defective graphene.

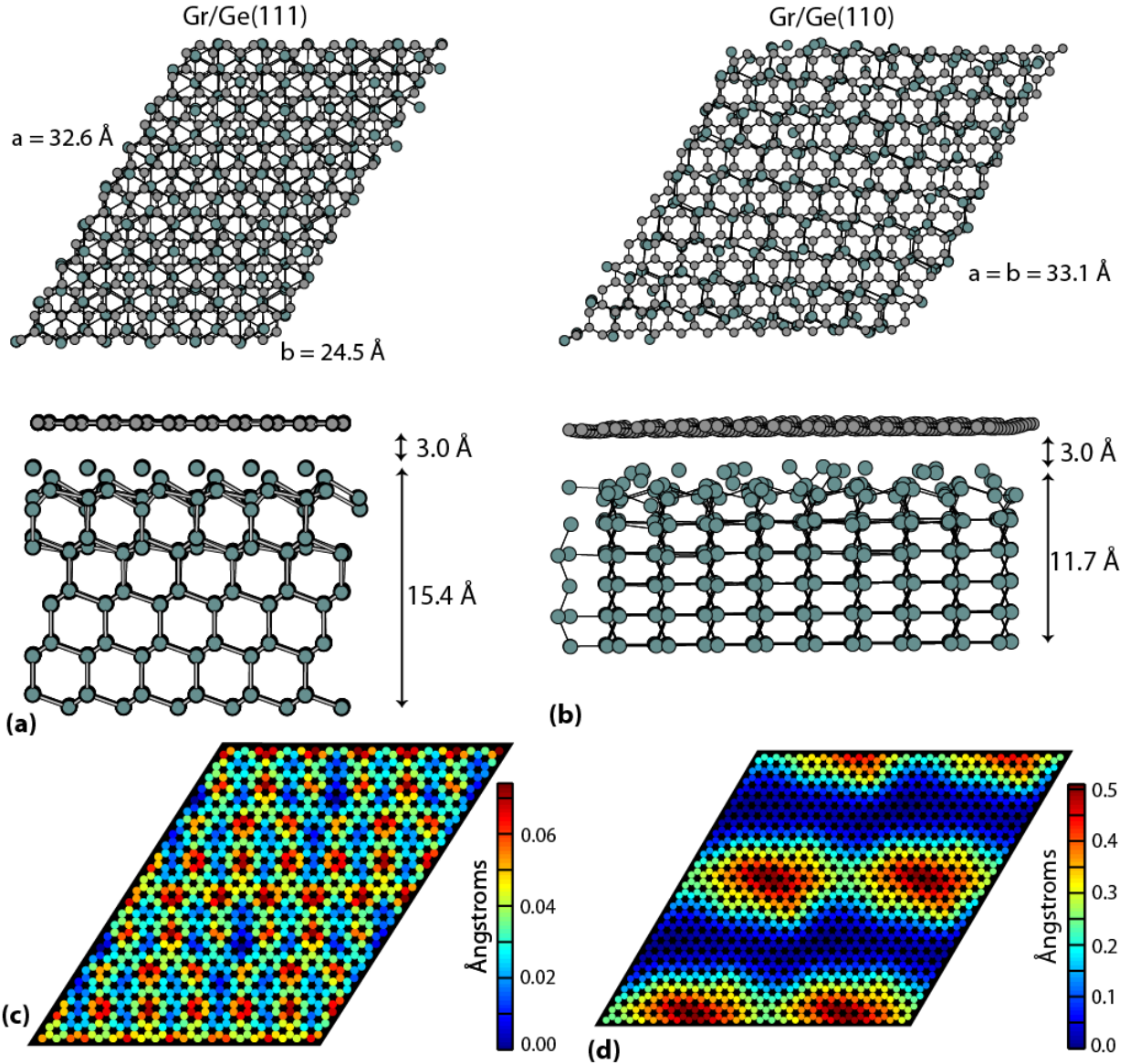

FIG. 5. Top and side views of DFT optimized graphene on (a) Ge(111)c(2x8) and (b) Ge(110)c(8x10) structures and relevant geometric parameters. Carbon atoms are colored grey and Ge atoms are green. (b) Top view of the graphene corrugations of four unit cells of the surfaces shown in (a). The colored scale bar indicates the distance between graphene and the Ge surface, where the distance is defined as the difference between a Ge adatom and the *closest* carbon atom.